\documentclass[]{article}
\usepackage{authblk}

\usepackage{subfigure, graphicx, amsmath, amssymb} 
\usepackage{booktabs}

\graphicspath{{images/}}

\begin{document}

\title{Learning Two-input Linear and Nonlinear Analog Functions with a Simple Chemical System}

\author[1]{Peter Banda\thanks{banda@pdx.edu}}
\author[2]{Christof Teuscher\thanks{teuscher@pdx.edu}}
\affil[1]{Department of Computer Science, Portland State University}
\affil[2]{Department of Electrical and Computer Engineering, Portland State University}


\date{1 April 2014}

\maketitle

\begin{abstract}
The current biochemical information processing systems behave in a pre-determined manner because all features are defined during the design phase. To make such unconventional computing systems reusable and programmable for biomedical applications, adaptation, learning, and self-modification based on external stimuli would be highly desirable. However, so far, it has been too challenging to implement these in wet chemistries.
In this paper we extend the chemical perceptron, a model previously proposed by the authors, to function as an analog instead of a binary system. The new analog asymmetric signal perceptron learns through feedback and supports Michaelis-Menten kinetics. The results show that our perceptron is able to learn linear and nonlinear (quadratic) functions of two inputs. To the best of our knowledge, it is the first simulated chemical system capable of doing so.
The small number of species and reactions and their simplicity allows for a mapping to an actual wet implementation using DNA-strand displacement or deoxyribozymes. Our results are an important step toward actual biochemical systems that can learn and adapt.
\end{abstract}

\vspace{10pt}


\section{Introduction}
\label{sec:introduction}

Biochemical information processing systems, which are crucial for emerging biomedical applications, cannot typically be programmed once built. After an \textit{in vitro} or \textit{in vivo} injection, the behavior, i.e., the program of such nano-scale chemical machines \cite{Wang2014} cannot be changed. That limits their applicability and re-usability. To address this limitation, future biochemical machinery should function not only in uniform, well-known lab settings but also in previously unknown environments. Such adaptive chemical systems would decide autonomously and learn new behaviors through reinforcements in response to external stimuli. We could imagine that in the future millions of molecular agents would help our immune system fight viruses, deliver medications \cite{LaVan2003}, or fix broken cells. Adaptive chemical systems may also simplify the manufacturing and design processes: instead of designing multiple systems with predefined functionality embedded in their species and reactions one could train and recycle a single adaptive machine for a desired functionality. 

Neural network theory \cite{hay09} inspired numerous chemical implementations \cite{bray95,mil99,kim04}, however, only the input-weight integration part of a single perceptron model \cite{ros58} was successfully mapped to chemistry. Learning (i.e., weight adaptation) was either not addressed or delegated to an external non-chemical system \cite{kim04,qian11} that calculated new weights values (i.e., chemical concentrations) to achieve a desired system behavior.

Our previous work \cite{Banda2013} introduced the first simulated artificial chemical system that can learn and adapt autonomously to feedback provided by a teacher. We coined the term {\em chemical perceptron} because the system qualitatively mimics a two-input binary perceptron. In a second step we aimed to simplify the model to make wet biochemical implementations feasible. We achieved that by employing the asymmetric representation of values and by using thresholding. The new \textit{asymmetric signal perceptron} (ASP) model \cite{Banda2014} requires less than a half of the reactions of its predecessors with comparable performance (i.e., $99.3-99.99\%$ success rates). The flip side of the more compact design is a reduced robustness to rate constant perturbations due to a lack of structural redundancy.

In real biomedical applications one is often required to distinguish subtle changes in concentrations with complex linear or nonlinear relations among species. Such behavior cannot easily be achieved with our previous binary perceptron models, thus, several improvements are necessary. In this paper we present a new {\em analog asymmetric signal perceptron} (AASP) with two inputs. We will refer to the original ASP as a binary ASP (BASP). The AASP model follows mass-action and Michaelis-Menten kinetics and learns through feedback from the environment. The design is modular and extensible to any number of inputs. We demonstrate that the AASP can learn various linear and nonlinear functions. For example, it is possible to learn to produce the average of two analog values. In combination with a chemical delay line \cite{Moles2014}, the AASP could also be used to predict time series.






\section{Chemical Reaction Network}
\label{sec:crn}

To model the AASP we employ the {\em chemical reaction network} (CRN) formalism. A CRN consists of a finite set of molecular species and reactions paired with rate constants \cite{espenson95}. CRN represents an unstructured macroscopic simulated chemistry, hence, the species labeled with symbols are not assigned a molecular structure yet. More importantly, since the reaction tank is assumed to be well-stirred, CRN lacks the notion of space. The state of the system does therefore not contain any spatial information and is effectively reduced to a vector of species concentrations. Without losing generality we treat a concentration as a dimensionless quantity. Depending on the required scale, a wet chemical implementation could use $mol \cdot L^{-1}$ ($M$) or $nmol \cdot L^{-1}$ ($nM$) with appropriate (scaled) rate constant units, such as $M \cdot s^{-1}$ or $M ^{-1} \cdot s^{-1}$, depending on the order of a reaction.

The reaction rate defines the speed of a reaction application prescribed by kinetic laws. The mass-action law \cite{espenson95} states that the rate of a reaction is proportional to the product of the concentrations of the reactants. For an irreversible reaction $aS_1 + bS_2 \rightarrow P$, the rate is given by
\begin{equation*}
 r = \frac{d[P]}{dt} = -\frac{1}{a}\frac{d[S_1]}{dt} = -\frac{1}{b}\frac{d[S_2]}{dt} = k[S_1]^a[S_2]^b,
\end{equation*}
where $k \in \mathbb{R}^+$ is a reaction rate constant, $a$ and $b$ are stoichiometric constants, $[S_1]$ and $[S_2]$ are concentrations of reactants (substrates) $S_1$ and $S_2$, and $[P]$ is a concentration of product $P$.

Michaelis-Menten enzyme kinetics \cite{cop02} describes the rate of a catalytic reaction $E + S \rightleftharpoons ES \rightarrow E + P$, where a substrate $S$ transforms to a product $P$ with a catalyst $E$, which increases the rate of a reaction without being altered. A species $ES$ is an intermediate enzyme-substrate binding. By assuming quasi-steady-state approximation, the rate is given by
\begin{equation*} 
r = \frac{d[P]}{dt} = \frac{k_{cat} [E][S]}{K_m + [S]},
\end{equation*}
where $k_{cat}, K_m \in \mathbb{R}^+$ are rate constants. By combining kinetic expressions for all species, we obtain a system of ODEs that we simulate using a $4^{th}$ order Runge-Kutta numerical integration with the temporal step $0.1$.
\section{Model}
\label{sec:model}

The AASP models a formal analog perceptron \cite{ros58} with two inputs $x_1$ and $x_2$, similar to an early type of artificial neuron \cite{hay09}. The perceptron is capable of simple learning and can be used as a building block of a feed-forward neural networks. Networks built from perceptrons have been shown to be  universal approximators \cite{hornik1989}.

In a CRN we represent each formal variable with one or several species. While the previous BASP models a perceptron with two inputs and a binary output produced by external or internal thresholding, the new AASP is analog and does not use thresholding. Instead of a binary yes/no answer, its output is analog, which requires much finer control over the weight convergence. As a consequence, the AASP consists of more species, namely 17 vs. 13, and more reactions, namely 18 vs. 16.


\begin{table}[h!]
  \centering
  \caption{(a) The AASP's species divided into groups according to their purpose and functional characteristics; (b) the AASP's reactions with the best rate constants found by the GA (see Section \ref{subsec:genetic-search}), rounded to four decimals. Groups $1-4$ implement the input-weight integrations, the rest implement learning. The catalytic reactions have two rates: $k_{cat}$ and $K_m$.}
  \scriptsize{
  \subtable{
  \begin{tabular}{l|r}
    \toprule
    Group Name & Species\\
	\hline
	Inputs                    & $X_1, X_2$ \\
	Output                    & $Y$ \\
	Weights                   & $W_0, W_1, W_2$ \\
	Target output             & $\hat{Y}$ \\
	Input (clock) signal      & $S_{in}$ \\
	Learning signal           & $S_L$ \\
	Input contributions       & $X_1Y, X_2Y, S_{in}Y$ \\
	Weight changers           & $W^\ominus, W^\oplus$ ,\\
	                          & $W_0^\ominus, W_1^\ominus, W_2^\ominus$ \\
      \bottomrule
	Total                     & $17$ \\
      \bottomrule
  \end{tabular}
  \label{tab:species}
  }
  \subtable{
    \begin{tabular}{c|r|r|r}
      \toprule
      Group & Reaction & Catalyst & Rates \\
      \hline
       1 & $ S_{in} + Y \rightarrow \lambda$          &           & .1800 \\
      \hline
       2 & $ S_{in} \rightarrow Y + S_{in}Y$          & $W_0$     & .5521, 2.5336 \\
      \hline
       3 & $ X_1 + Y \rightarrow \lambda$             &           & .3905\\
         & $ X_2 + Y \rightarrow \lambda$             &           & \\
      \hline
       4 & $ X_1 \rightarrow Y + X_1Y$                & $W_1$     & .4358, 0.1227\\
         & $ X_2 \rightarrow Y + X_2Y$                & $W_2$     & \\
      \bottomrule
       5 & $ \hat{Y} \rightarrow W^\oplus$            &           & .1884 \\
      \hline
       6 & $ Y \rightarrow W^\ominus$                 & $S_L$     & .1155, 1.9613\\
      \hline
       7 & $ Y + \hat{Y} \rightarrow \lambda$         &           & 1.0000\\
      \hline
       8 & $ W^\ominus \rightarrow W_0\ominus$        & $S_{in}Y$ & 0.600, 1.6697\\
      \hline
       9 & $ W_0 + W_0\ominus \rightarrow \lambda$    &           & .2642\\
      \hline
       10 & $ W^\oplus \rightarrow W_0$               & $S_{in}Y$ & .5023, 2.9078\\
      \hline
       11 & $ W^\ominus \rightarrow W_1^\ominus$      & $X_1Y$    & .1889, 1.6788\\
          & $ W^\ominus \rightarrow W_2^\ominus$      & $X_2Y$    & \\
      \hline
       12 & $ W_1 + W_1^\ominus \rightarrow \lambda$  &           & .2416\\
          & $ W_2 + W_2^\ominus \rightarrow \lambda$  &           & \\
      \hline
       13 & $ W^\oplus \rightarrow W_1$               & $X_1Y$    & .2744, 5.0000\\
          & $ W^\oplus \rightarrow W_2$               & $X_2Y$    & \\
      \bottomrule
	Total                     & \multicolumn{3}{c}{$18$} \\
      \bottomrule
    \end{tabular}
    \label{tab:reactions}
  }
  }
\end{table}

\subsection{Input-Weight Integration}
\label{subsec:input-weight-integration}
A formal perceptron integrates the inputs $\mathbf{x}$ with the weights $\mathbf{w}$ linearly as $\Sigma_{i=0}^n w_i \cdot x_i$, where the weight $w_0$, a bias, always contributes to an output because its associated input $x_0 = 1$. An activation function $\varphi$, such as a hyperbolic tangent or signum, then processes the dot product to produce the output $y$.

The reactions carrying out the chemical input-weight integration are structurally the same as in the BASP. The only difference is an addition of the partial input-weight contribution species, which are, however, required for learning only, and will be explained in Section \ref{subsec:input-weight-integration}. The AASP models a two-input perceptron where the output calculation is reduced to $y = \varphi (w_0 + w_1 x_1 + x_2 w_2)$. The concentration of input species $X_1$ and $X_2$ corresponds to the formal inputs $x_1$ and $x_2$, and the species $Y$ to the output $y$. A clock (input) signal $S_{in}$ is always provided along the regular input $X_1$ and $X_2$, since it serves as the constant-one coefficient (or the constant input $x_0 = 1$) of the bias weight $w_0$.

The AASP represents the weights by three species $W_1, W_2$, and $W_0$. As opposed to the formal model, the input-weight integration is nonlinear and based on an annihilatory version of the asymmetric representation of the values and the addition/subtraction operation as introduced in \cite{Banda2014}. Since the concentration cannot be negative, we cannot map a signed real variable directly to the concentration of a single species. The weights require both positive and negative values, otherwise we could cover only functions that are strictly additive. The asymmetric representation uses a single species $E$ that catalyzes a transformation of substrate $S$ to a product $P$ ($S \xrightarrow{E} P$) and competes against an annihilation of the substrate and the product $S + P \rightarrow \lambda$. For a given threshold concentration of the product we can determine the associated catalyst threshold, so all concentrations of catalyst $[E]_0$ to the left of this threshold represent negative numbers while all concentrations to the right represent positive numbers. The final product concentration $[P]_\infty$ is monotonically increasing and asymptotically reaches the initial concentration of the substrate $[S]_0$ for $[E]_0 \rightarrow \infty$.

\begin{figure}[ptbh]
  \centering
  \subfigure[input-weight integration] {
    \includegraphics[scale=.43]{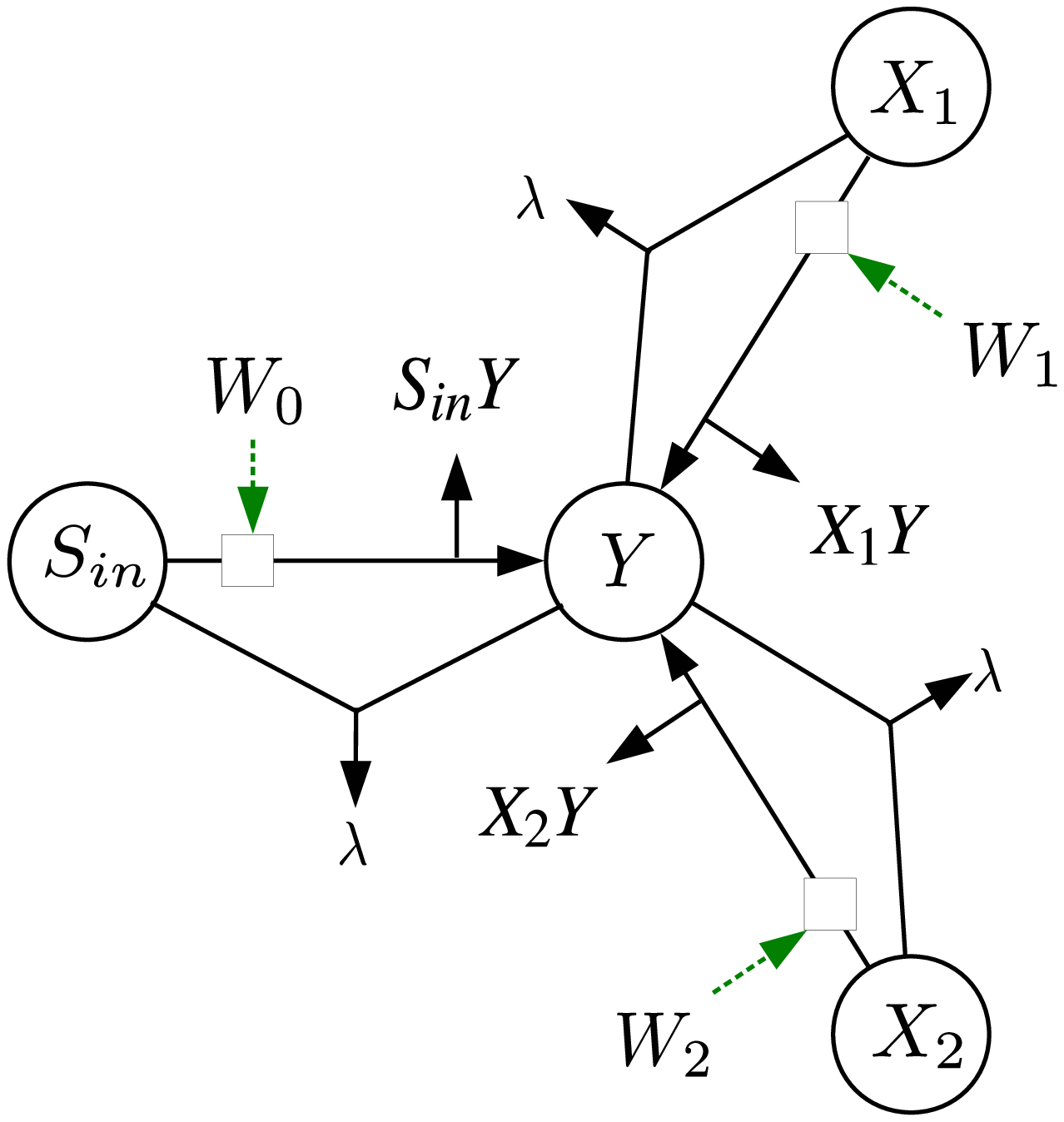}
    \label{fig:input-weight_integration}
  }
  \subfigure[output comparison] {
     \raisebox{20mm}{\includegraphics[scale=.43]{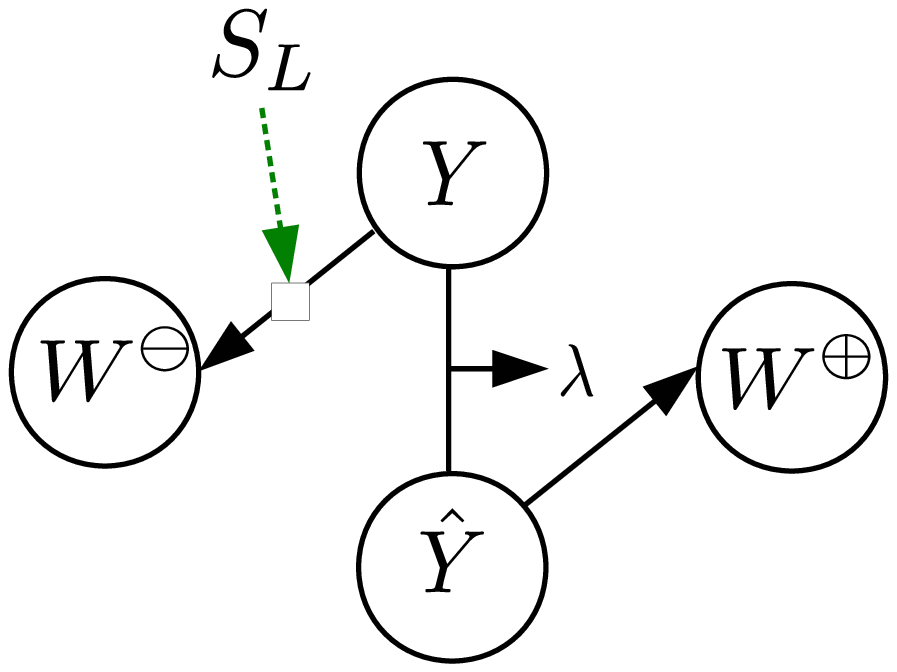}}
     \label{fig:learning-core}
  }
  \subfigure[positive adaptation] {
     \raisebox{14mm}{\includegraphics[scale=.43]{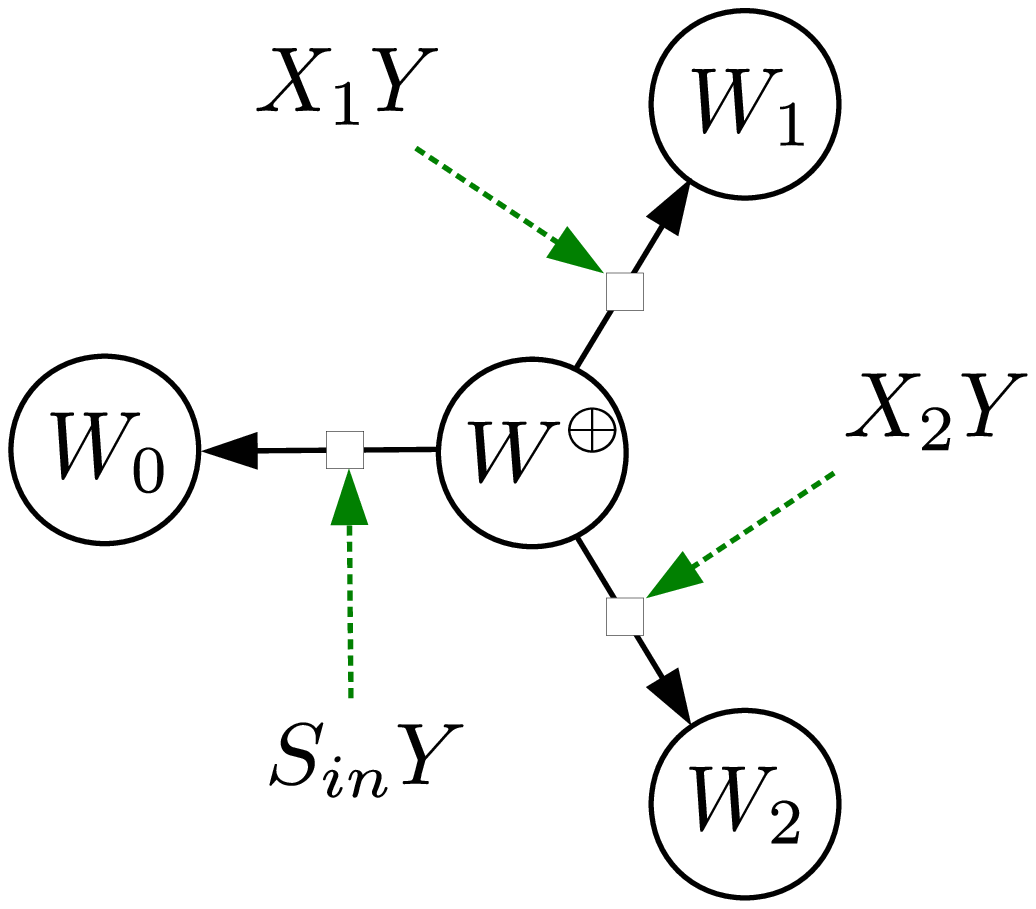}}
     \label{fig:learning-positive}
  }
  \subfigure[negative adaptation] {
     \includegraphics[scale=.43]{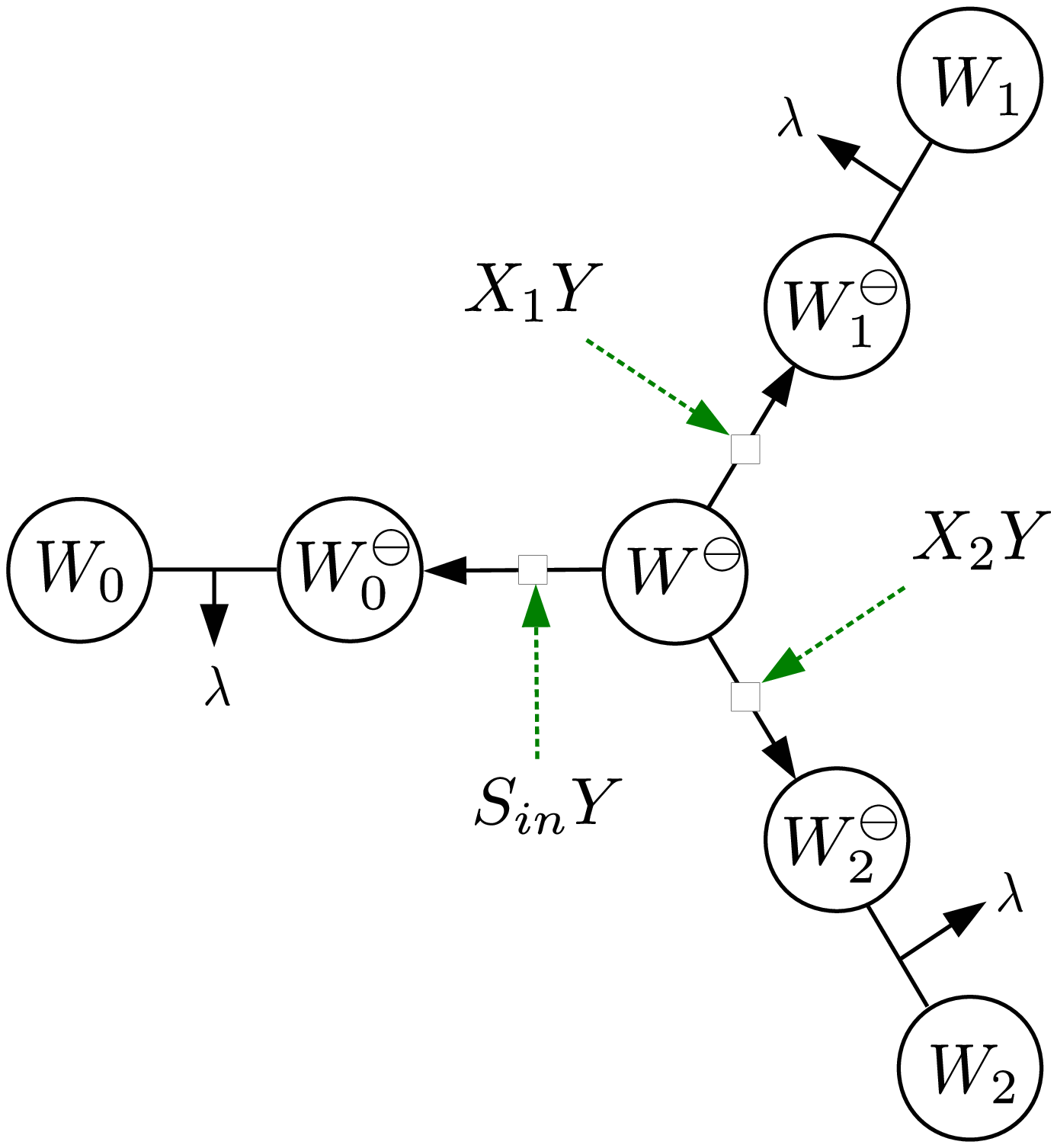}
     \label{fig:learning-negative}
  }
  \caption{(a) The AASP's reactions performing input-weight integration. Similarly to the BASP, cross-weight competition is achieved by the annihilation of the inputs $S_{in}, X_1, X_2$ with the output $Y$, an asymmetric strategy for representation of real values and subtraction. (b-d) the AASP's reactions responsible for learning. They are decomposed into three parts: (b) comparison of the output $Y$ with the target-output $\hat{Y}$, determining whether weights should be incremented ($W^{\oplus}$ species) or decremented ($W^{\ominus}$ species), and (c-d) positive and negative adaptation of the weights $W_0, W_1$, and $W_2$, which is proportional to the part of the output they produced $S_{in}Y, X_1Y$, and $X_2Y$ respectively. Nodes represent species, solid lines are reactions, dashed lines are catalysts, and $\lambda$ stands for no or inert species.}
  \label{fig:AASP-full}
\end{figure}

Using the asymmetric comparison primitives, we map the AASP's weights to catalysts ($E$), the inputs to substrates ($S$), and the output to product ($P$) and obtain $6$ reactions as shown in Figure \ref{fig:input-weight_integration} and Table \ref{tab:reactions}, groups $1-4$. Each weight species races with its substrate's annihilation but also with other weights. Since the output $Y$ is shared, this effectively implements a nonlinear input-weight integration. Note that by replacing annihilation with a decay of input species, we would end up having three independent races with additive contributions instead of one global race. An alternative symmetric representation embedded in the previously reported \textit{weight-loop perceptron} and the \textit{weight-race perceptron} \cite{Banda2013} encodes the values by two complementary species, one for the positive and one for the negative domain. We opt for the asymmetric approach because it reduces the number of reactions by half compared to the symmetric one.


Because of the complexity of the underlying ODEs, no closed formula for the output concentration exists and theoretical conclusions are very limited. Although we cannot analyze the input-weight integration dynamics quantitatively, we can still describe the qualitative behavior and constraints. The weight concentration represents formally both positive and negative values, so the weights together with annihilatory reactions can act as both catalysts and inhibitors. More specifically a low weight concentration, which strengthens its input-specific annihilation, could impose a negative pressure on a different weight branch. Hence, we interpret a weight that contributes to the output less than its input consumes as negative. In an extreme case, when the weight concentration is zero, its branch would consume the same amount of output as its input injected. The relation between the concentration of weights and the final output $[Y]_\infty$ has a sigmoidal shape with the limit $[X_1]_0 + [X_2]_0 + [S_{in}]_0$ reaching for all weights $[W_i] \rightarrow \infty$. Clearly the output concentration cannot exceed all the inputs provided.

\begin{figure}[h!]
  \centering
  \subfigure[$x_1=.2,x_2=.2$] {
     \includegraphics[scale=.44]{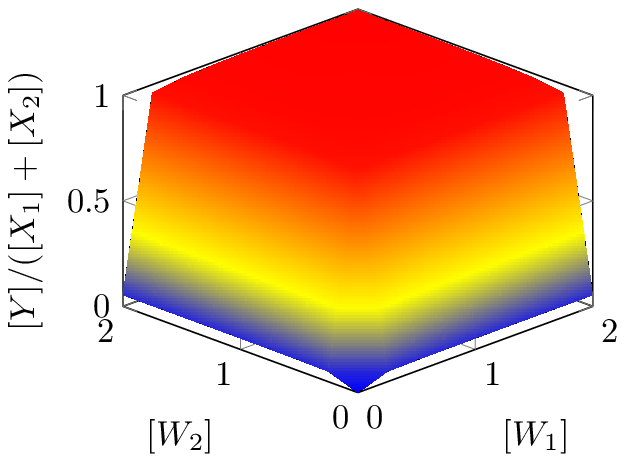}
     \label{fig:asymmetric_comparisons_mm_anih_anih_0_2_x1_0_2_x2_0_2}
  }
  \subfigure[{$x_1=.6,x_2=.6$}] {
     \includegraphics[scale=.44]{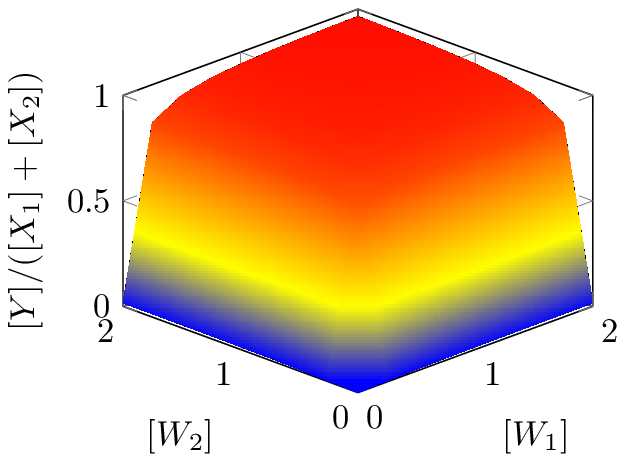}
     \label{fig:asymmetric_comparisons_mm_anih_anih_0_2_x1_0_6_x2_0_6}
  }
  \subfigure[{$x_1=1,x_2=1$}] {
     \includegraphics[scale=.44]{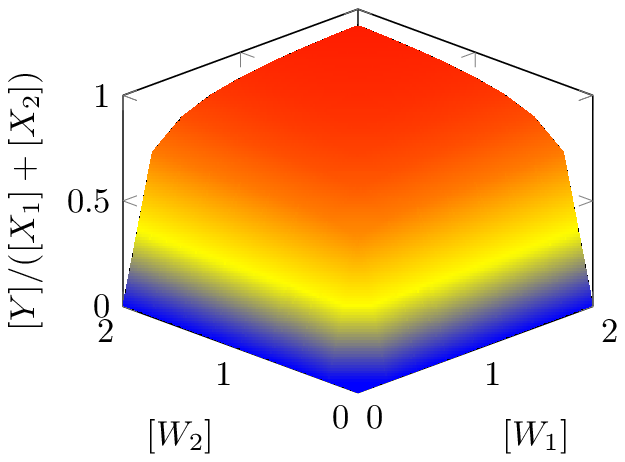}
     \label{fig:asymmetric_comparisons_mm_anih_anih_0_2_x1_1_0_x2_1_0}
  }
  \subfigure[$x_1=.2,x_2=.8$] {
     \includegraphics[scale=.44]{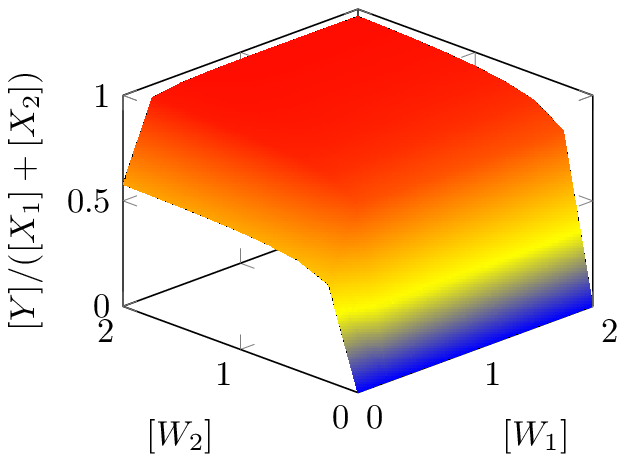}
     \label{fig:asymmetric_comparisons_mm_anih_anih_0_2_x1_0_2_x2_0_8}
  }
  \subfigure[$x_1=.2,x_2=.2$] {
     \includegraphics[scale=.44]{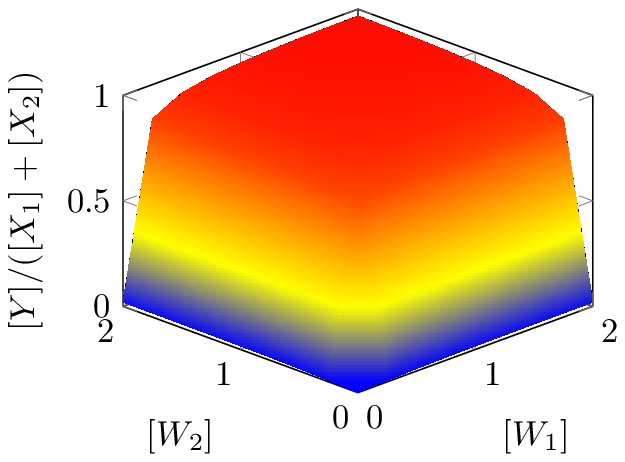}
     \label{fig:asymmetric_comparisons_mm_anih_anih_1_0_x1_0_2_x2_0_2}
  }
  \subfigure[$x_1=.6,x_2=.6$] {
     \includegraphics[scale=.44]{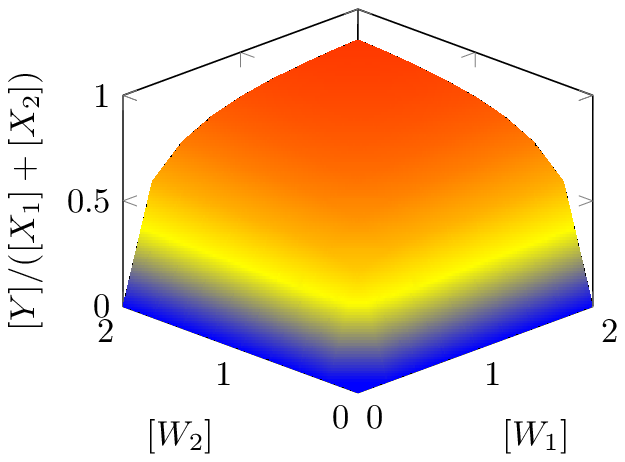}
     \label{fig:asymmetric_comparisons_mm_anih_anih_1_0_x1_0_6_x2_0_6}
  }
  \subfigure[$x_1=1,x_2=1$] {
     \includegraphics[scale=.44]{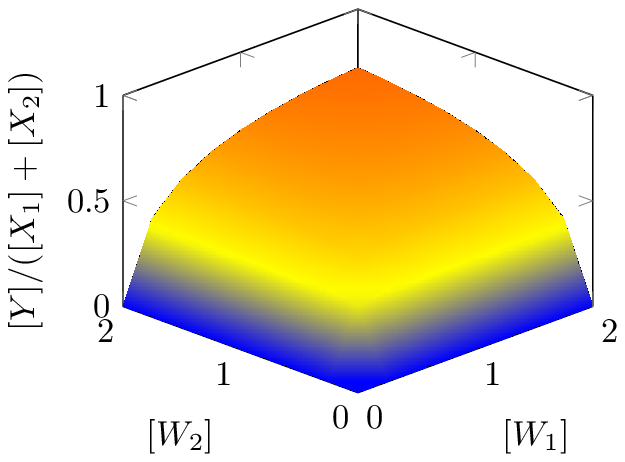}
     \label{fig:asymmetric_comparisons_mm_anih_anih_1_0_x1_1_0_x2_1_0}
  }
  \subfigure[$x_1=.2,x_2=.8$] {
     \includegraphics[scale=.44]{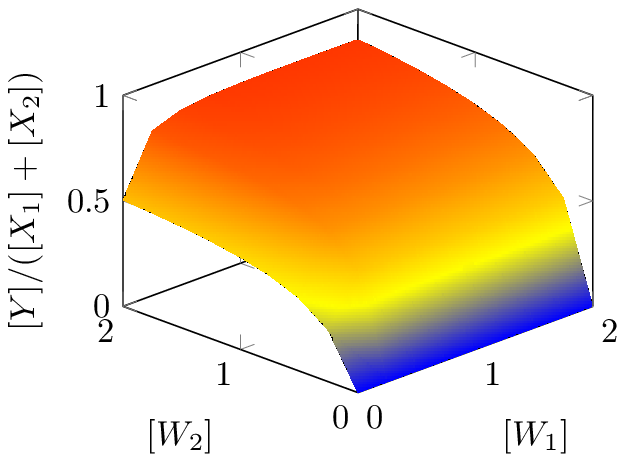}
     \label{fig:asymmetric_comparisons_mm_anih_anih_1_0_x1_0_2_x2_0_8}
  }
  \caption{The relation between the weight concentrations $[W_1]$ and $[W_2]$ and the final output concentration $[Y]_\infty$ normalized by $[X_1]_0 + [X_2]_0$ for the input-weight integration (excluding the bias $W_0$ part) showing various inputs. The rate constant of annihilatory reactions $X_i + Y \rightarrow \lambda, i \in \{1,2\}$ is $k = 0.2$ in the top and $k = 1$ in the bottom row.}
  \label{fig:asymmetric-comparisons}
\end{figure}

Figure \ref{fig:asymmetric-comparisons} shows the relation between the concentration of weight $W_1$ and weight $W_2$ and the final output concentration. For simplicity the bias processing part is not considered ($[S_{in}] = 0$), so we keep only two branches of the input-weight integration triangle. Note that in the plots the concentration of weights span the interval $0$ to $2$ because in our simulations we draw the weights uniformly from the interval $(0.5,1.5)$. On the z-axis we plotted the ratio of the output concentration $[Y]$ to $[X_1]_0 + [X_2]_0$. For learning to work we want the gradient of the output surface to be responsive to changes in the weight concentrations. Therefore, we restrict the range of possible outputs so it is neither too close to the maximal output, where the surface is effectively constant, nor too close to zero, where the surface is too steep and even a very small perturbation of the weight concentration would dramatically change the output. Note that we optimized the AASP's rate constants to obtain an optimal weight-output surface by genetic algorithms (discussed in Section ~\ref{subsec:genetic-search}).

\subsection{Learning}
\label{subsec:learning}

In the previous BASP model, learning reinforced the adaptation of weights by a penalty signal, whose presence indicated that the output was incorrect. Since the output is analog in the new AASP model, a simple penalty signal is not sufficient anymore. We therefore replaced the reinforcement learning by classical supervised learning \cite{roj96}. Formally, the adaptation of a weight $w_i$ for the training sample $(\mathbf{x}, \hat{y})$, where $\hat{y}$ is a target output, and $\mathbf{x}$ a input vector, is defined as $\bigtriangleup w_i = \alpha (\hat{y} - y(t))x_i$, where $\alpha \in (0,1]$ is the learning rate. The AASP's, similarly to the input-weight integration, does not implement the formal $\bigtriangleup w_i$ adaptation precisely, rather, it follows the relation qualitatively. 

The learning is triggered by an injection of the target output $\hat{Y}$ provided some time after the injection of the input species. The part presented in Figure \ref{fig:learning-core} compares the output $Y$ and the target output $\hat{Y}$ by annihilation. Intuitively a leftover of the regular output $Y$ implies that the next time the AASP faces the same input, it must produce less output, and therefore it needs to decrease the weights by producing a negative weight changer $W^\ominus$ from $Y$. In the opposite case, the AASP needs to increase the weights, hence $\hat{Y}$ transforms to a positive weight changer $W^\oplus$. Since the AASP can produce output also without learning, just by the input-weight integration, we need to guard the reaction $Y \rightarrow W^\ominus$ by a learning signal $S_L$, which is injected with the target output and removed afterwards. To prevent creation of erroneous or premature weight changers, the annihilation $Y + \hat{Y} \rightarrow \lambda$ must be very rapid. Note that the difference between the actual output $Y$ and the desired output $\hat{Y}$, materializing in the total concentration of weight changers $W^{\oplus}$ and $W^{\ominus}$, must not be greater that the required weight adaptation, otherwise the weights would diverge. The learning rate $\alpha$ is therefore effectively incorporated in the concentration of $W^{\oplus}$ and $W^{\ominus}$.

In the formal perceptron, the adaptation of a weight $w_i$ is proportional to the current input $x_i$. Originally, the BASP distinguished which weights to adapt by a residual concentration of inputs $X_1$ and $X_2$. Because the inputs as well as an adaptation decision were binary, we cared only about whether some of the unprocessed input were still left, but not about its precise concentration. Thus, an injection of the penalty signal could not happen too soon, neither too late. Because the AASP's learning needs more information, the input-weight integration introduced three additional species, namely the partial input-weight contributions $X_1Y$, $X_2Y$, $S_{in}Y$, which are produced alongside the regular output $Y$. A decision which weights to update based on the input-weight contributions could be made even after the input-weight integration is finished. That allows to postpone an injection of the target output $\hat{Y}$ and the learning signal $S_L$.

Let us now cover a positive adaptation as shown in Figure \ref{fig:learning-positive}, where the total amount of $W^\oplus$ is distributed among participating weights. The input contribution species $X_1Y, X_2Y, S_{in}Y$ race over the substrate $W^\oplus$ by catalyzing the reactions $W^\oplus \rightarrow W_i, i \in \{0,1,2\}$. Note that the traditional weight adaptation formula takes into count solely the input value, so here we depart further from the formal perceptron and have the combination of input and weights compete over $W^\oplus$. Since larger weights produce more output they get adapted more. In addition, once a weight reaches zero, it will not be recoverable.

The negative adaptation presented in Figure \ref{fig:learning-negative} is analogous to the positive one, but this time the input-weight contributions race over $W^\ominus$ and produce intermediates $W_0^\ominus, W_1^\ominus, W_2^\ominus$, which annihilate with the weights. Again, because the magnitude of a weight update depends on the weight itself, this feedback loop protects the weight from falling too low and reaching zero (i.e., a point of no return). This is beneficial because as opposed to the formal perceptron, a weight value (concentration) cannot be physically negative.

To implement the entire learning algorithm, the AASP requires $12$ reactions as presented in Table \ref{tab:reactions}, groups $5-13$.




\subsection{Genetic Search}
\label{subsec:genetic-search}

Since a manual trial-and-error setting of the rate constants would be very time-consuming, we optimize the rate constants by a standard genetic algorithm (GA). Possible solutions are encoded on chromosomes as vectors of rate constants, which undergo cross-over and mutation. We use elite selection with elite size $20$, $100$ chromosomes per generation, shuffle cross-over, per-bit mutation, and a generation limit of $50$. The fitness of a chromosome defined as the RNMSE reflects how well the AASP with the given rate constants (encoded in the chromosome) learns the target functions $k_1x_1 + k_2x_2 + k_0$, $k_1x_1$, and $k_2x_2$. The fitness of a single chromosome is then calculated as the average over $300$ runs for each function. We included the $k_1x_1$ and $k_2x_2$ tasks to force the AASP to utilize and distinguish both inputs $x_1$ and $x_2$. Otherwise the GA would have a higher tendency to opt for a greedy statistical approach where only the weight $W_0$ (mean) might be utilized.

\section{Performance}
\label{sec:performance}

We demonstrate the learning capabilities of the AASP on $6$ linear and nonlinear target functions as shown in Table \ref{tab:target-functions}. During each learning iteration we inject inputs $X_1$ and $X_2$ with concentrations drawn from the interval $(0.2, 1)$ and set the bias input $S_{in}$ concentration to $0.5$. We chose the target functions carefully, such that the output concentration is always in a safe region, i.e., far from the minimal (zero) and the maximal output concentration $[S_{in}]_0 + [X_1]_0 + [X_2]_0$. We then inject the target output $\hat{Y}$ with the learning signal $S_L$ $50$ steps after the input, which is sufficient to allow the input-weight integration to proceed.

\begin{figure}[t!]
  \centering
  \includegraphics[width=.7\textwidth]{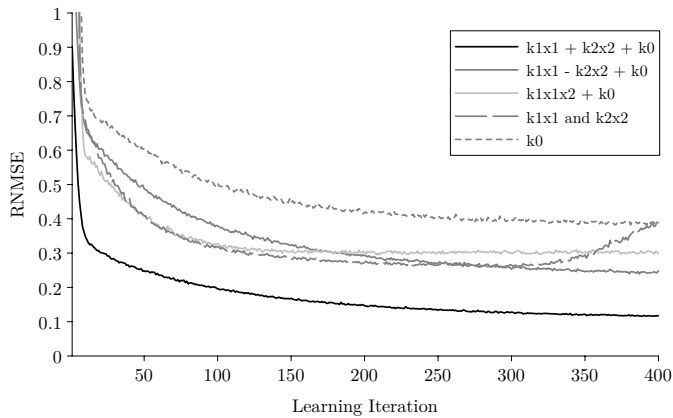}
  \caption{RNMSE for $6$ linear and nonlinear functions over $400$ learning iterations.}
  \label{fig:learning_performance}
\end{figure}

For each function family we calculated the AASP's performance over $10,000$ simulation runs, where each run consists of $400$ training iterations. We define performance as the root normalized mean square error (RNMSE)
\begin{equation*}
\text{RNMSE} = \sqrt[]{\frac{\langle(y - \hat{y})^2 \rangle}{\sigma_{\hat{y}}^2}}.
\end{equation*}
A RNMSE of $1$ means chance level. The AASP's RNMSE settles down to the range $(0.117, 0.0.388)$ (see Figure \ref{fig:learning_performance}), which implies that it learns and generalizes all target functions sufficiently.
When we include only the functions that utilize both inputs $x_1$ and $x_2$, as well as the bias, i.e., the scenario the AASP was primarily designed for, RNMSE drops to the range $(0.117, 0.298)$.
Note that we do not distinguish between the training and testing set. During each iteration we draw the inputs with the target output for a given function independently. 

\begin{table}[b!]
  \centering
  \caption{Target functions with uniform constant $k_1, k_2, k_0$ intervals.}
  \begin{tabular}{l|r|r|r}
    \toprule
    \multicolumn{1}{c}{$\hat{y}$} & \multicolumn{1}{c}{$k_1$} & \multicolumn{1}{c}{$k_2$} & \multicolumn{1}{c}{$k_0$} \\
	\hline
	$k_1 x_1 + k_2 x_2 + k_0$  & $(0.2, 0.8)$ & $(0.2, 0.8)$ & $(0.1, 0.4)$ \\
	$k_1 x_1 - k_2 x_2 + k_0$  & $(0.2, 0.8)$ & $(0.0, 0.3)$ & $(0.4, 0.7)$ \\
	$k_1 x_1$                  & $(0.2, 0.8)$ & $-$          & $-$ \\
	$k_2 x_2$                  & $-$          & $(0.2, 0.8)$ & $-$ \\
	$k_1 x_1 x_2 + k_0$        & $(0.2, 0.8)$ & $-$          & $0.25$ \\
	$k_0$                      & $-$          & $-$          & $(0.1, 0.4)$ \\
    \bottomrule
  \end{tabular}
  \label{tab:target-functions}
\end{table}

\begin{figure}[h!]
  \centering
  \subfigure[$\hat{y} = k_1x_1 + k_2x_2 + k_0$] {
     \includegraphics[width=.5\textwidth]{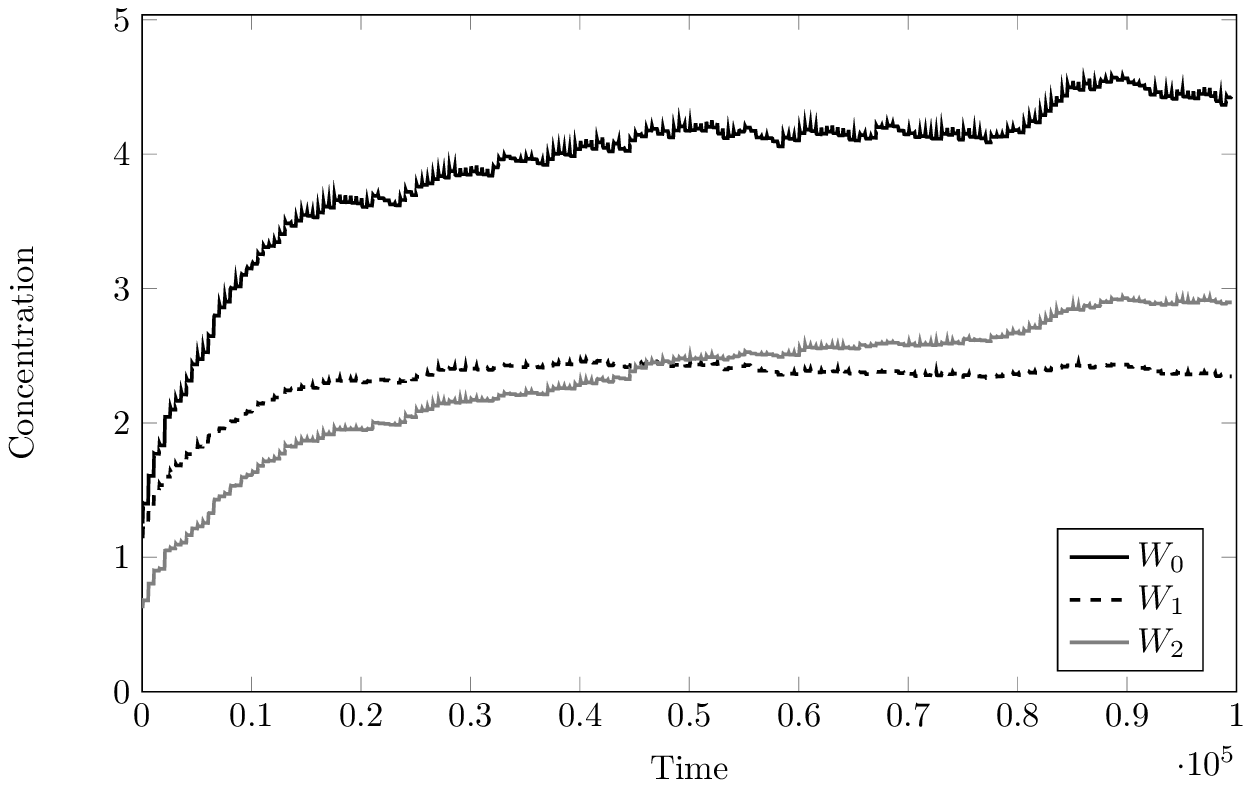}
     \includegraphics[width=.5\textwidth]{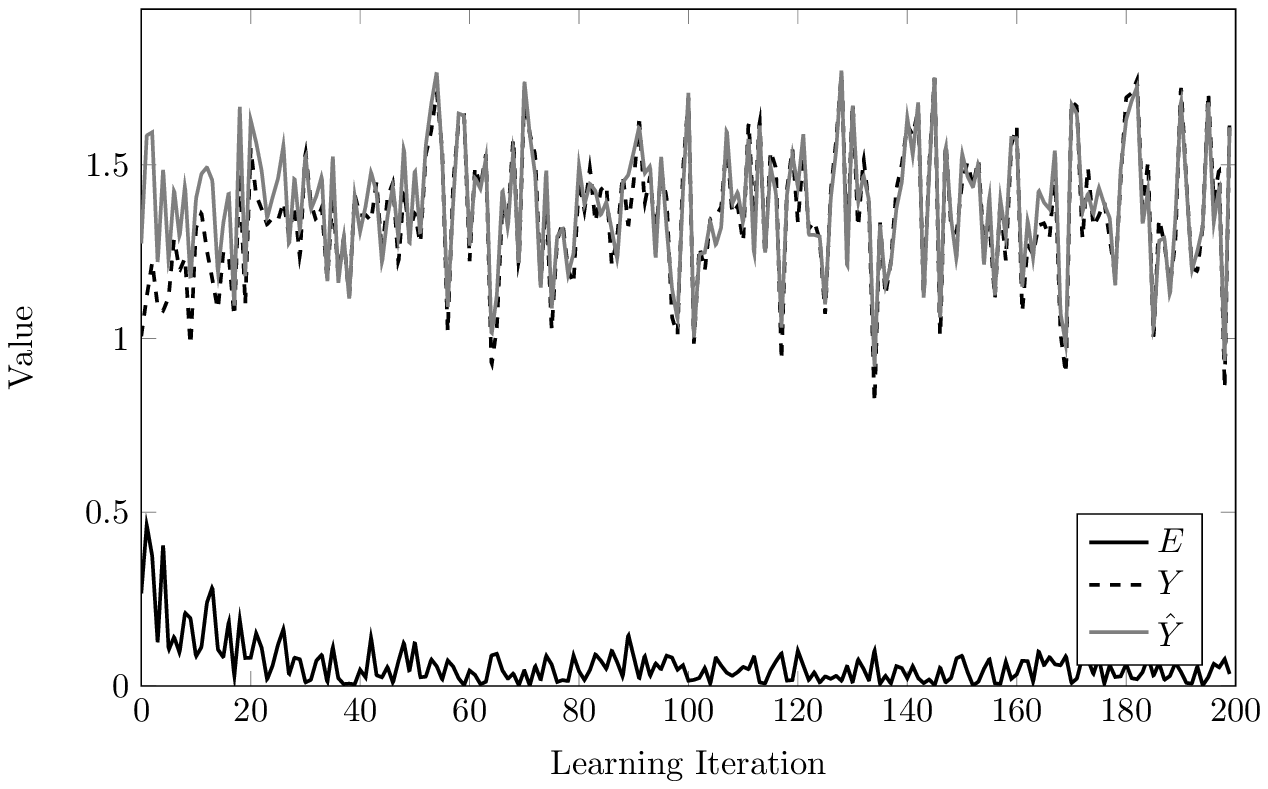}
     \label{fig:weights-output-rvp_v41_k1x1+k2x2+k0}
  }
  \subfigure[$\hat{y} = k_0$] {
     \includegraphics[width=.5\textwidth]{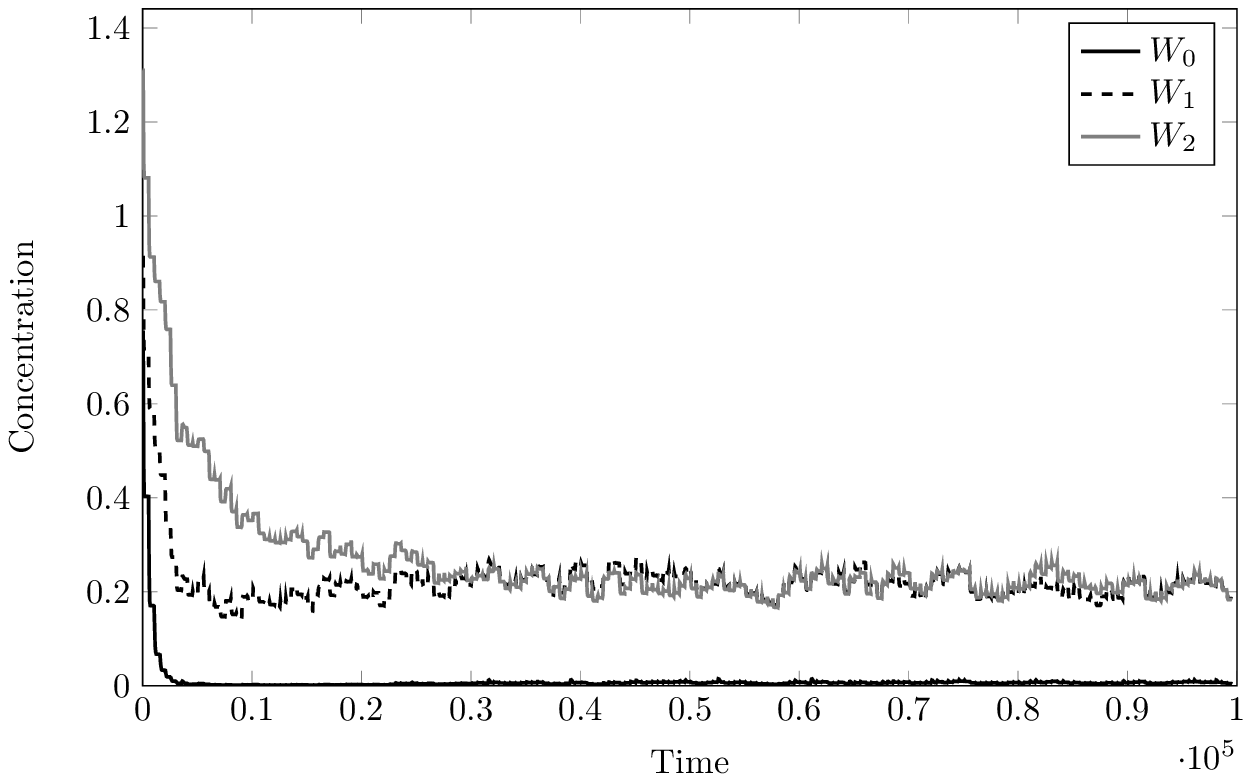}
     \includegraphics[width=.5\textwidth]{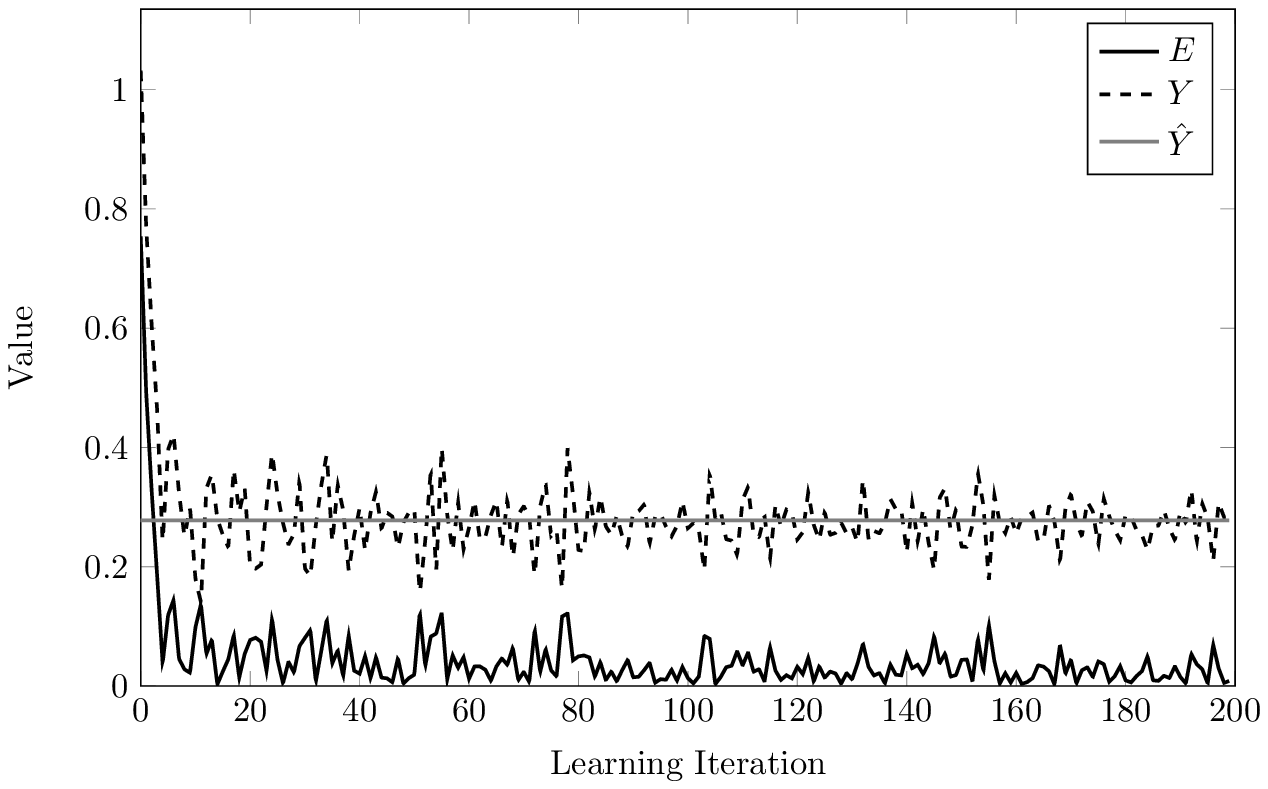}
     \label{fig:weights-output-rvp_v41_k}
  }
  \subfigure[$\hat{y} = k_1x_1x_2 + k_0$] {
     \includegraphics[width=.5\textwidth]{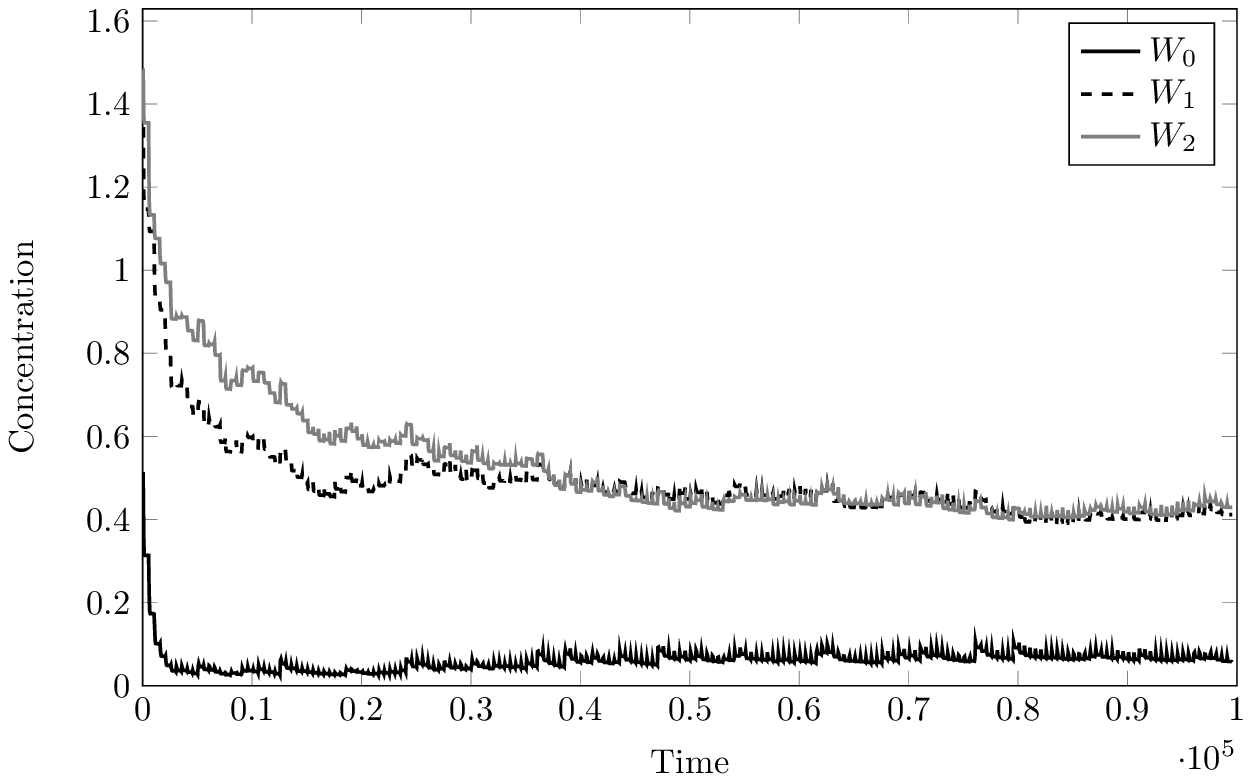}
     \includegraphics[width=.5\textwidth]{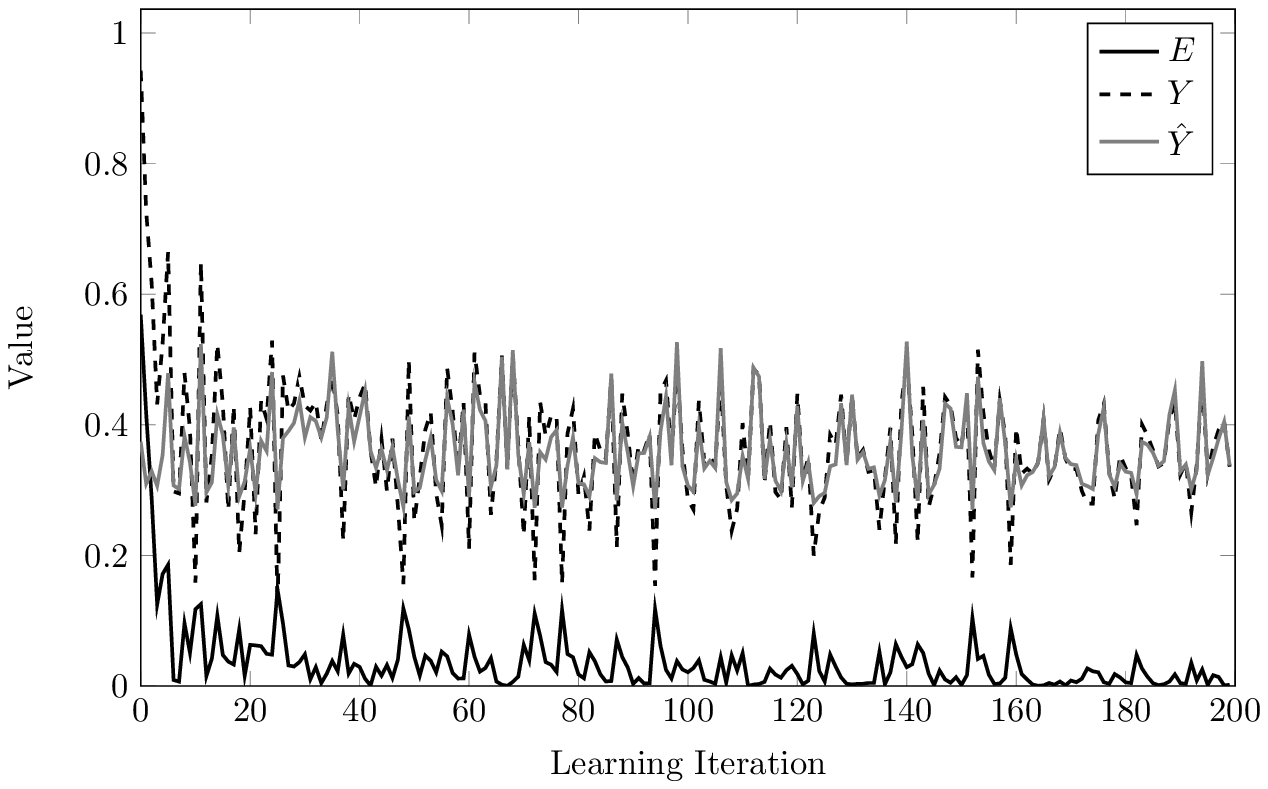}
     \label{fig:weigts-output-rvp_v41_kx1x2+0_25}
  }
  \caption{AASP learning examples for selected functions. The left column shows concentration traces of the weights, the right column the filtered output, the target output, and the absolute error.}
  \label{fig:learning_examples}
\end{figure}

Among all the functions, $k_1x_1 + k_2x_2 + k$ is the easiest (RNMSE of $0.117$) and the constant function $k_0$ the most difficult (RNMSE of $0.388$) one. The function $k_0$ is even more difficult than the nonlinear function $k_1x_1x_2 + k_0$ (RNMSE of $0.298$). Compared to the formal perceptron, the constant function does not reach zero RNMSE because the AASP cannot fully eliminate the contribution (or consumption) of the $X_1$ and $X_2$ input-weight branches. The formal perceptron could simply discard both inputs and adjust only the bias weight, however, the AASP's weights $W_1$ and $W_2$ with zero concentration would effectively act as inhibitors, thus consuming a part of the output produced by the bias. On the other hand, a nonlinear $k_1x_1x_2 + k_0$ function with fairly low RNMSE would be impossible to calculate for the formal perceptron. Therefore it is an open question what function classes can be learned by the AASP. Note that for the nonlinear function we set $k_0 = 0.25$, which does not increase the variance, i.e., only the nonlinear part counts toward the error. Figure \ref{fig:learning_examples} shows the weight concentration traces as well as the output, the target output, and the absolute error for selected functions.


\section{Conclusion}
\label{sec:conclusion}

In this paper we extended our chemical asymmetric design introduced for the asymmetric signal perceptron to an analog scenario. We demonstrated that our new AASP model can successfully learn several linear and nonlinear two-input functions. The AASP follows Michaelis-Menten and mass-action kinetics, and learns through feedback provided as a desired output.

In related work, Lakin et al. \cite{LakinMR:towblm} designed and simulated a system based on enzymatic chemistry, capable of learning linear functions of the form $k_1x_1 + k_2x_2$. Compared to the AASP, the system lacks cross-weight competition, meaning the weights could not formally represent negative numbers, and so the system could model only strictly additive functions with $k_1, k_2 \geq 0$. Besides regular inputs $x_1$ and $x_2$ the AASP utilizes also the bias (constant shift), hence it can model linear functions of a more general form $k_1 x_1 + k_2 x_2 + k_0$ as well as nonlinear (quadratic) functions of the form $k x_1 x_2 + k_0$, where $k_1, k_2, k_0 \in \mathbb{R}$. The AASP uses $18$ reactions, however, by excluding the bias ($k_0$) part, it would need just $13$ as opposed to $27$ reactions employed in Lakin's system. On the other hand, Lakin's system targets a specific wet implementation based on deoxyribozyme chemistry, so the higher number of reactions is justifiable. Last but not least, we evaluated the performance more precisely over $10,000$ instead of $10$ trials.

Because the number of species and reactions employed is fairly low, a wet chemical implementation is plausible. More precisely, we suggest that the AASP could be mapped to catalytic DNA chemistry ~\cite{stoj03,Liu2009} by having each catalysis carried out by deoxyribozime-substrate cleavage. The most problematic part for this mapping would be the feedback reactions, where each of three enzymes, $X_1Y,X_2Y,$ and $S_{in}Y$, catalyzes two reactions, which is non-trivial to implement in practice. To address that we would need to introduce two variants of a feedback enzyme $X_iY^{\oplus}$ ($S_{in}Y^{\oplus}$) and $X_iY^{\ominus}$ ($S_{in}Y^{\ominus}$) to separate these two reaction pathways. Alternatively we could obtain a wet chemical implementation of the AASP automatically by Soloveichik's transformation \cite{sol10}, which compiles mass-action driven CRN to DNA-strand displacement reactions \cite{zha11}. That would produce a chemical circuit with around $80$ different DNA strands, which is in the range of other state-of-the-art DNA circuits.

As opposed to our previous designs using simple binary signals, the AASP allows to adapt to precise concentration levels. By integrating the AASP with a chemical delay line as proposed in \cite{Moles2014}, we could also tackle time-series prediction. Consequently, chemical systems would be able monitor concentrations of selected molecular species and respond if a severe event, defined as a linear or nonlinear temporal concentration pattern, occurs. Such a system would be highly relevant where the quantity or type of the drug required could be adjusted in real-time with complex relations among species, e.g., produced by cancer cells.


\section*{Acknowledgment}
This material is based upon work supported by the National Science Foundation under grant no. 1028120.

\bibliography{bibliography}
\bibliographystyle{splncs}

\end{document}